# Phase Coherent Transport of Charges in Graphene Quantum Billiard

F. Miao, S. Wijeratne, U. Coskun, Y. Zhang and C.N. Lau[*]
Department of Physics and Astronomy, University of California, Riverside, Riverside, CA 92521

**Abstract**
We experimentally investigate electrical transport properties of graphene, which is a two dimensional (2D) conductor with relativistic energy dispersion relation. By investigating single- and bi-layer graphene devices with different aspect ratios, we confirm experimentally that the minimum conductivity in wide and short graphene strips approaches the universal value of $4e^2/\pi h$. At low temperatures, quantum interference of multiply-reflected waves of electrons and holes in graphene give rise to periodic conductance oscillations with bias and gate voltages. Thus graphene acts as a quantum billiard, a 2D ballistic, phase coherent electron system with long phase coherence length that exceeds 5 μm. Additional features in differential conductance emerge when graphene is coupled to superconducting electrodes. We observe proximity-induced enhanced conductance at low bias, and conductance dips at energy scales far above the superconducting gap of the electrodes. The latter provides preliminary evidence for a novel superconducting material that consists of graphene coated with metallic atoms.

[*] To whom correspondence should be addressed. Email: lau@physics.ucr.edu

Graphene, a two-dimensional (2D) a honey-comb lattice of carbon atoms, exhibits rather unusual energy dispersion relations – the low-lying electrons in single layer graphene behave like massless relativistic Dirac fermions with vanishing density of states at the Dirac point, and bilayer's band structure resembles that of a zero band gap semiconductor (Fig. 1a). Since recent experimental isolation and measurement of graphene[1-3], it has attracted tremendous attention, as the special band structures in single and bi-layer graphenes yield novel aspects to the physics of two-dimensional electron systems. The Dirac spectrum in graphene is predicted to give rise to a number of phenomena, such as quantum spin hall effects[4-6], enhanced Coulomb interaction[7-11], and suppression of weak localization[12, 13]. Technologically, graphene is an attractive material for nanoscale electronics engineering. As a two-dimensional (2D) relative of carbon nanotubes, it manifests high mobility, extraordinary thermal conductivity and atomic perfection; but in contrast to nanotubes, traditional lithographic techniques can potentially be employed for tailoring of transport properties and device synthesis[14].In the past two years, much progress has been made on theoretical understanding of the novel electronic properties that may emerge in graphene. In contrast, experimental measurements of graphene have been relatively scarce.

Here we demonstrate experimentally that single layer (SLG) and bi-layer (BLG) graphene can act as a quantum billiard, i.e. a 2D ballistic system where scattering only occur at boundaries, with a phase coherence length that exceeds 5 μm. The ballistic transport for charge carriers, coupled with phase coherent multiple-reflection at the electrodes, give rise to quantum interference of waves of the charge carriers, thus realizing a quantum resonator for electrons and holes. Moreover, our experimental results shed light on one of the central controversies in graphene: the exact value of conductivity at the Dirac point in the ballistic regime. Contrary to one's naïve expectation that scattering from impurities would reduce measured value of conductivity, the experimentally determined value[2] appears to be *larger* than theoretical predictions[15-20]. We address the controversy by demonstrating that the minimum conductivity of a wide graphene strip approaches the theoretical value of *$4e^2/\pi h$*; for narrower strips, our data agrees with recent theoretical prediction with no free parameters[17].

The long phase coherence length suggests that Cooper pairs may be readily transmitted through graphene. In the second part of the paper, we couple graphene with superconducting electrodes, and explore how ballistic transport is modified by the presence of superconducting order. We find that the quantum interference patterns persist, with additional low-bias conductance peaks that arise from multiple Andreev reflections at the superconductor-graphene interface. This may open the possibility to explore the physics of an Andreev billiard, which is an important tool for understanding quantum chaos[21]. Additionally, we observed a dip in conductance at energy scales far above *Δ/e,* providing preliminary evidence for a novel superconducting material based on graphene coated with palladium adatoms, where the electron-electron attraction is predicted to be mediated not by phonons, but by plasmons[22].

Single- and bi-layer graphene devices are fabricated following procedures outlined in ref. [1]. Pieces of graphene are mechanically exfoliated and transferred to silicon wafers, which are degenerately doped to serve as back gates for the devices. We identify single and bi-layer graphenes by color inference in optical microscope, and use atomic force microscopy (AFM) to ensure the uniformity of the layer. The electrodes are patterned by electron beam lithography, followed by deposition of a 10-nm layer of palladium(Pd) and 70 nm of aluminum(Al). The electrodes are separated by distances ranging from 100 nm to 400 nm. SLG devices are fabricated in the Hall bar geometry (Fig. 1a), and bilayer devices in both Hall bar and two-terminal geometry (Fig. 1b). The device parameters are listed in Table 1. The devices are measured in our Helium-III refrigerator with a base temperature of 260mK. The electrical lines in the refrigerator are fitted with three stages of filters, similar to those described in ref. [23], to eliminate high frequency noise and ensure low electron temperatures.

For those devices with Hall bar geometry, quantum hall measurements were carried out. The devices were current-biased at 100 nA with an applied magnetic field $B$=8T, while transverse and longitudinal voltage differences were measured as a function of gate voltage $V_g$. Modulating $V_g$ tunes both the sign and the magnitude of the charge density $n$ of graphene. A simple estimate of $n$ can be obtained by considering the capacitance between the graphene and the back gate, given by $n = \frac{\varepsilon V_g}{te}$, where $e$ is the electron charge, $\varepsilon$=4.5 is the dielectric constant of silicon oxide, and $t$ is the thickness of the silicon oxide layer. Here $V_g$ is the gate voltage measured from the charge neutrality point $V_{CN}$, which is experimentally determined to range from -1.2V to -64V (see Fig. 2a). In our devices, $n/V_g \sim 10^{11}$ cm$^{-2}$V$^{-1}$. The quantum hall results are displayed in Fig. 1c, where the transverse conductivity $\sigma_{xy}$ are plotted as a function of gate voltage $V_g$. Clear quantization at half-integer values of $G_Q=4e^2/h\sim 160\mu S$ are observed at both positive and negative gate voltages, confirming the anomalous integer quantum hall effect in massless Dirac fermion systems. This unequivocally establishes our selection of SLG.

We first investigate electrical transport properties of the devices at 1.5 K, which is above the superconducting transition temperature $T_c$ of bulk Al (~ 1.2 K). As shown in Fig. 2a, the device conductance is high at large negative and positive gate voltages, corresponding to regimes where graphene is highly hole- and electron-doped, respectively. At the charge neutrality point, the density of state vanishes and the device conductance reaches a minimum $G_{min}$, which ranges from 150 to 350 μS for our devices. A distinct feature of a ballistic 2D Dirac fermion system is that the minimum *conductivity* $\sigma_{min}$ has been predicted to be a universal value $4e^2/\pi h$ [15-20]. Experimentally, an earlier study[2] indeed found a universal value for both single and bi-layer graphene, albeit at a different value of $4e^2/h$. This factor of $\pi$ discrepancy between theory and experiment has since been a topic of much debate[15-18, 24]. To gain insight into this controversy, we plot measured values of $\sigma_{min}$ (in units of $4e^2/\pi h$) as a function of the aspect ratio $W/L$, where data from SLG and BLG are represented by red squares and blue triangles, respectively (Fig. 2b). [25] We can see that when $W/L<2$, $\sigma_{min}$ is ~ 2 – 4, consistent with results from the previous study; however, for the device with an aspect ratio of 8, $\sigma_{min}$ is close to 1. Our results are in agreement with recent theoretical work[17]: even in ballistic regime, $\sigma_{min}$ depends on the graphene's aspect ratio and as well as microscopic details of the edges, and only approaches the universal value of $4e^2/\pi h$ for wide and short graphene strips with $W/L>4$.

We can quantitatively compare our results with the theoretical prediction. At the Dirac point, the conductivity of a graphene strip with arm-chair edges is given by[17]

$$\sigma_{min} = G_{min}\frac{L}{W} = \frac{4e^2}{h}\frac{L}{W}\sum_{n=0}^{N-1}T_n, \qquad (1)$$

where $T_n$ is the transmission probability of the nth conductance channel,

$$T_n = \frac{1}{\cosh^2(\pi nL/W)} \qquad (2)$$

Using (1), (2) and $N$=100, we calculate $\sigma_{min}$ as a function of $W/L$. The result is plotted in Fig. 2b (dotted line), in excellent agreement with experimental data. It is also worth noting that the bilayer devices appear to fall on a different curve, as one would expect from theoretical predictions [26]. Thus, our experimental results provide evidence for the universal $\sigma_{min}=4e^2/\pi h$ in ballistic SLG observed at large aspect ratios.

To further investigate ballistic charge transport in graphene, we measure the differential conductance ($dI/dV$) of the devices at 260mK as functions of source-drain bias and gate voltage. A small magnetic field (~40mT) is applied perpendicular to the graphene plane in order to suppress superconductivity in the aluminum electrodes. The data, shown as 2D color plots, (Fig. 3a and 3b), display striking patterns of criss-crossing lines, indicating periodic conductance oscillation with both bias and gate voltages. Similar patterns are observed in both single and bi-layer devices, and in both hole- and electron-doped regimes. Such patterns arise from quantum interference of multiply-reflected paths of electron (and hole) waves between two partially transmitting electrodes, whereas the charges' wavelength is tuned by bias or gate voltage. We note that similar interference patterns was observed in carbon nanotubes[27], with a characteristic energy scale $E_0 = \frac{hv_F}{2L}$, where $v_F$ ~10$^6$m/s is the Fermi velocity of graphene. For $L$=100 nm, $E_0$~ 20 meV. The length $2L$ also yields the minimum length over which electrons remain phase coherent. In graphene devices, the smallest energy scale we have unambiguously identified is ~ 0.7 meV, corresponding to a phase coherent length of more than 5 μm. Thus, in our devices graphene acts as an electronic analogue of Fabry-Perot resonant cavity, supporting ballistic, phase coherent transport of electrons and holes over relatively large distances.

Despite the remarkable similarity between the data from graphene and nanotubes, the two differ in one crucial aspect – in nanotubes typically one finds only a single frequency characterized by $E_0$, whereas we generally observe several frequencies with different characteristic energy scales. An example is shown in Fig. 3a, where a large period and a small period are outlined by dotted and dashed lines, respectively. In general, we found that the characteristic energy scale in the conductance plot for each device increases with decreasing source-drain separation, but we are not able to establish a quantitative relationship. This distinction of nanotubes from graphene arises from the key difference between the two systems – the former is a 1D device with well-defined electron paths, whereas the latter is 2D and the electron trajectories can be quite complicated, as various electronic resonator modes (and hence energy scales) can be excited[28, 29]. The hall bar geometry of the devices further complicates the situation, since the resonator is open and electrons have finite probability of escaping the cavity. Thus, graphene acts as a quantum billiard, where the electron trajectories may be chaotic and non-integrable, and further experimental investigation and theoretical modeling will be necessary for quantitative understanding of the system.

The long phase coherence length of charges in graphene suggests a possible experimental realization of Andreev billiard, that is, a quantum billard coupled to superconducting contacts. Such a device is also a novel Josephson junction, where the electron system in the normal metal is not a Fermi Liquid, but Dirac fermions. These junctions are predicted to exhibit a number of novel phenomena, such as specular Andreev reflection[30, 31], novel propagating modes of Andreev electrons in the normal metal channel[32], and oscillation of transmission probability with the barrier width[33]. To explore this possibility, we perform measurement at zero magnetic field with superconducting Al electrodes. As shown in Fig. 4a, the interference pattern persists, with some additional features. The most prominent feature is a bright, horizontal band around zero-bias, indicating enhanced conductance that persists through both the resonance and anti-resonance of the interference patterns. Comparing with data taken at 50mT, conductance of devices coupled to superconductors is enhanced by as much as 25% for bias below certain characteristic voltage $V_1$, which ranges from 105 μV to 200 μV (Fig. 4b and 4c). As we expect superconducting proximity effect to give rise to conductance enhancement for $V_{bias} \leq 2\Delta/e$, where $\Delta$ is the superconducting gap, we take $2\Delta \sim eV_1$ at the graphene-electrode interface, which is comparable to the expected value of $\Delta_{bulk} \approx 180$ μeV for bulk aluminum. The reduced gap for some devices can be attributed to the presence an unusually thick (10 nm) adhesion layer of Pd that partially suppresses aluminum's superconducting order at the graphene-electrode interface.

A detailed examination of the low bias region reveals a single peak at zero bias, and two additional peaks at $V_2$, where $V_2 \sim \pm 30$ - 60μV for different devices (Fig. 4b, 4c and 4e). The zero bias anomaly has been observed in nanotubes[34-36], and has been attributed to either a dissipative quasiparticles current[37], or to a manifestation of supercurrent [36]. The two side peaks, occurring at voltages below $\Delta$, are the so-called subgap structures, and generally attributed to multiple Andreev reflections at normal metal-superconductor interfaces[38, 39]. To further elucidate origins of these peaks, we study their behavior in small magnetic field. In Fig. 4d, the differential conductance is plotted as functions of bias (vertical) and magnetic field (horizontal) for Device 7, and the enhanced conductance peaks appear as the three red bands near zero bias. We can see that the positions of these peaks move to lower bias with increasing magnetic field, and decreases to zero at B=19mT, which is comparable to the critical field $H_c$ for bulk aluminum, ~ 10mT. This provides further evidence that the low bias peaks in $dI/dV$ arise from proximity effect due to the superconducting Al electrodes. This constitutes the first observation of multiple Andreev reflection peaks through *single* layer graphene (a very recent article reported similar findings for few-layer graphenes [40]). We note that even though enhanced conductance due to proximity effect is observed, we did not observe supercurrent through the devices (although in a very recent report supercurrent through graphene was reported [41]). This may be attributed to intrinsic effects, such as suppression of localization and breaking of time-reversal symmetry in graphene[13, 42], or to extrinsic effects, such as insufficiently filtered lines. Further experimental efforts will be conducted to clarify this issue.

Finally, we focus on the conductance dips at ~1-3mV occurs at energies far above superconducting gap. Similar to the low-bias conductance peaks, they are independent of gate voltage. Such dips in differential conductance are found in every device with transparent contacts, although at different bias values that range from ~0.8 to 2.5 mV. We also measured a few devices with resistance above 20kΩ, and no conductance dip was observed. We note that similar above-gap features was observed in quantum wells coupled to superconductors, and was attributed to suppression of excess current due to re-captured Andreev holes back-scattered from electrodes[43]. However, such processes are not allowed in our geometry. Here we pursue an alternative explanation: these dips at high bias indicate the presence of a superconductor with an energy gap several times that of Al. Indeed, a recent theory predicts plasmon-mediated superconductivity in single layer graphene coated with a dilute layer of metallic adatoms[22]. Thus, this small dip at ~1-2mV may arise from burgeoning superconductivity in the region where the Pd/Al electrodes contact graphene. The variation in the voltages at which the dips occur may be attributed to the uncontrolled interface of Al/Pd/graphene. A tantalizing clue is provided by the dips' dependence on magnetic field (Fig. 3e). Although the low-bias peaks are suppressed at B=19mT, these above-gap conductance dips (appear as the white and blue bands) persist until ~32mT, while following the BCS dependence for classical superconductors. This suggests the presence of a superconductor with $T_c$ higher than Al. For quantitative comparison, we plot the energy gap of a superconductor as a function of magnetic field, using[44]

$$\Delta(H) = \Delta(0)\sqrt{1 - \frac{H^2}{H_c^2(0)}} \qquad (3)$$

Taking $H_c(0)$=31.5mT and $\Delta(0)$=0.7meV directly from the data, the resulting curve is superimposed on the Fig. 3d (dotted line), in excellent agreement with data. Thus, the magnetic field dependence of these conductance dips is consistent with that of a superconductor with a relatively large energy gap. While not conclusive evidence for a novel plasmon-mediated superconductor, our data provide motivation for further experimental efforts to investigate this intriguing phenomenon.

**Acknowledgement**
We thank Shan-Wen Tsai, Marc Bockrath, Leonid Pryadko, Michael Tinkham and Chandra Varma for stimulating discussions.

**Table I. List of device parameters.**
(SL: single layer; BL: Bi-layer. Hall and 2-terminal refers to device geometry.)

| Device | Geometry | $W$ (nm) | $L$ (nm) | $R_{min}$ (kΩ) |
|---|---|---|---|---|
| Device 1 | SL, Hall | 640 | 385 | 5.28 |
| Device 2 | SL, Hall | 360 | 385 | 6.48 |
| Device 3 | SL, Hall | 360 | 270 | 6.09 |
| Device 4 | SL, Hall | 320 | 460 | 6.81 |
| Device 5 | SL, Hall | 840 | 110 | 3.2 |
| Device 5 | BL, 2-terminal | 1000 | 380 | 2.94 |
| Device 6 | BL, 2-terminal | 850 | 300 | 2.86 |
| Device 7 | BL, Hall | 560 | 280 | 3.1 |

**Fig. 1.** (a).Schematics of the energy dispersion relations for single-layer (left) and bi-layer (right) graphene. (b) Scanning electron microscope image of a single layer graphene devices in Hall bar geometry. (c) AFM image of a device with two-terminal geometry. The device shown here has a measured thickness of 3.5 nm, indicating the presence of several atomic layers. (d). Quantum hall effect observed in single-layer graphene devices at B=8T and 260mK. As the gate voltage is modulated, the Hall resistivity $\sigma_{xy}$ are quantized at half-integer values of $4e^2/h$.

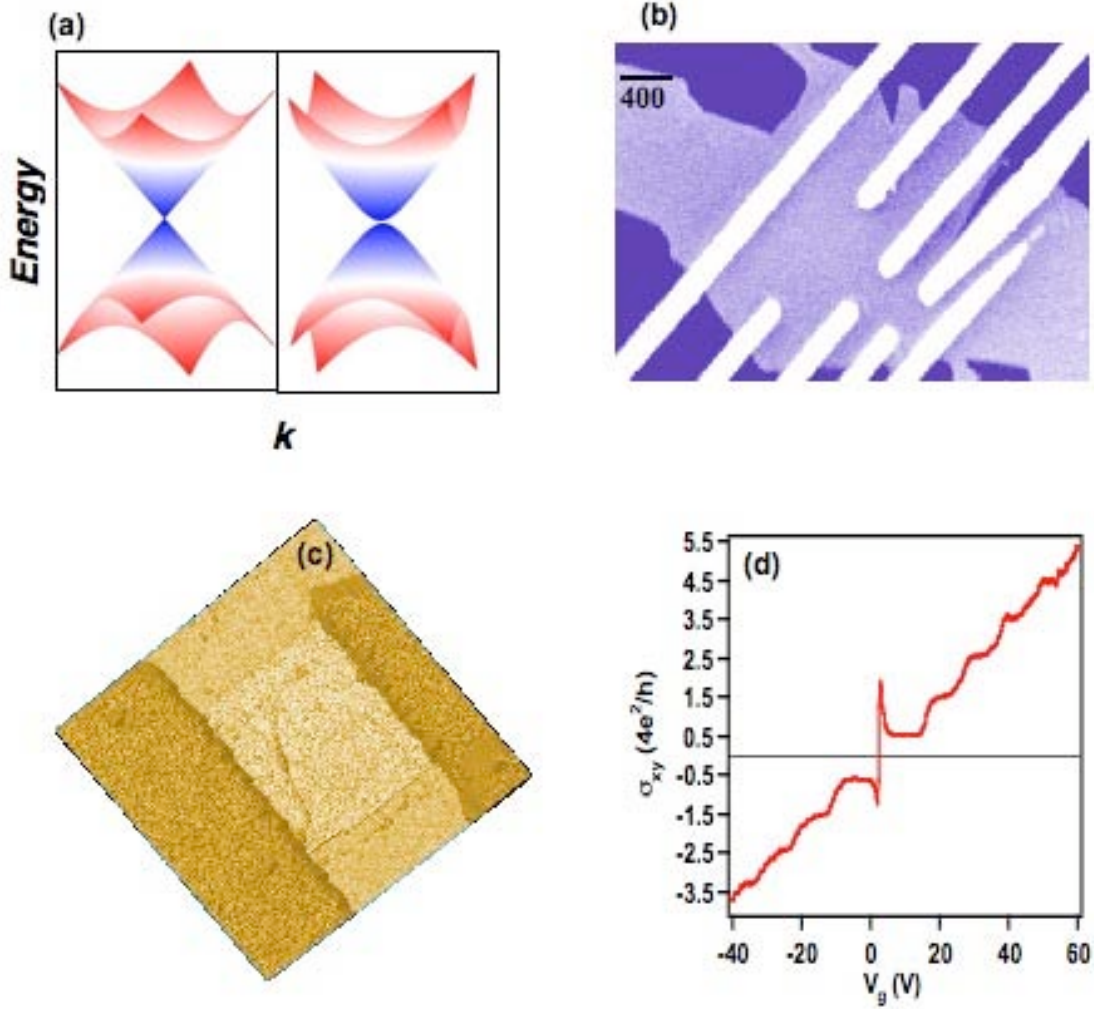

**Fig. 2.** (a) Conductance as a function of gate voltage for single and bi-layer devices (lower and upper panel, respectively), demonstrating the ambipolar behavior of the devices. (b) Minimum conductivity of the devices (in units of $4e^2/\pi h$) as a function of device aspect ratio $W/L$. Red Squares: single layer devices. Blue Triangles: Bi-layer devices. The dotted line is theoretical prediction calculated from Eq. (1) and (2) *with no free parameters*.

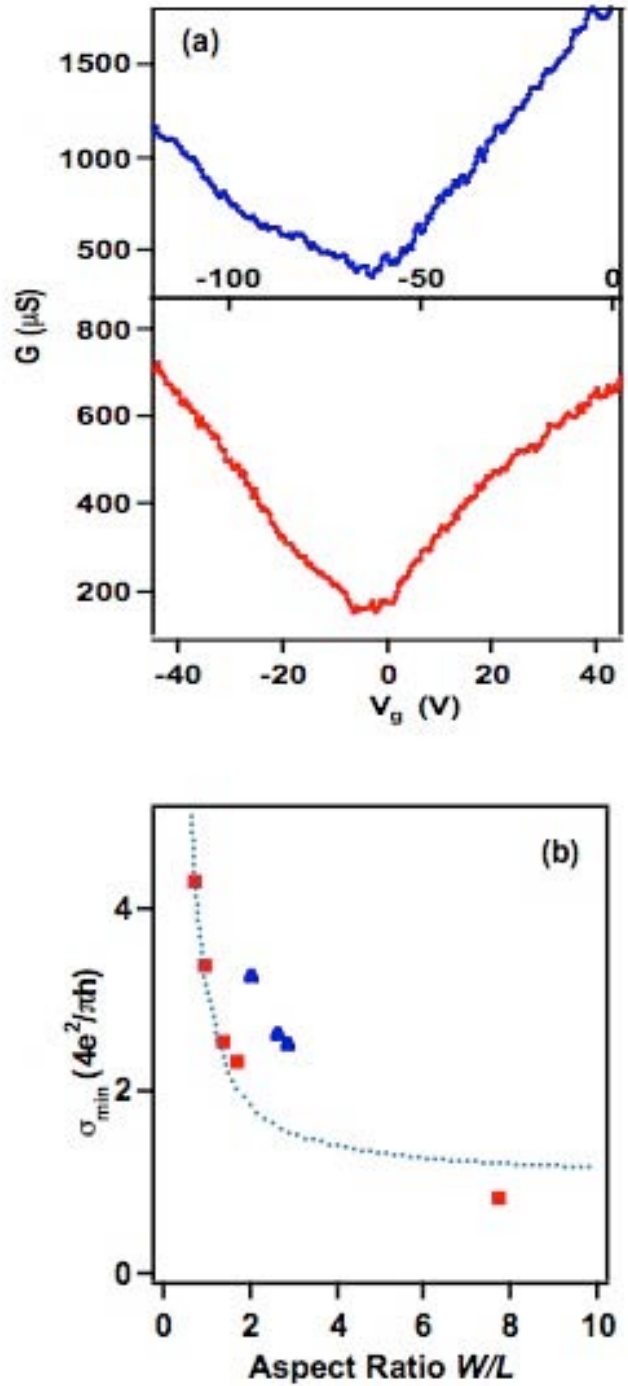

**Fig. 3**. (a) (b) Differential conductance (color scale, in μS) as functions of gate voltage (horizontal axis) and source-drain bias (vertial axis) for Device 1 and Device 7 at 260mK. 40mT is applied to suppress superconductivity in the aluminum electrodes. The dotted and dashed lines in 4a are guides to the eye indicating the periodic oscillation of conductance. (c) Line trace through 3b at $V_{bias}$=1.3mV.

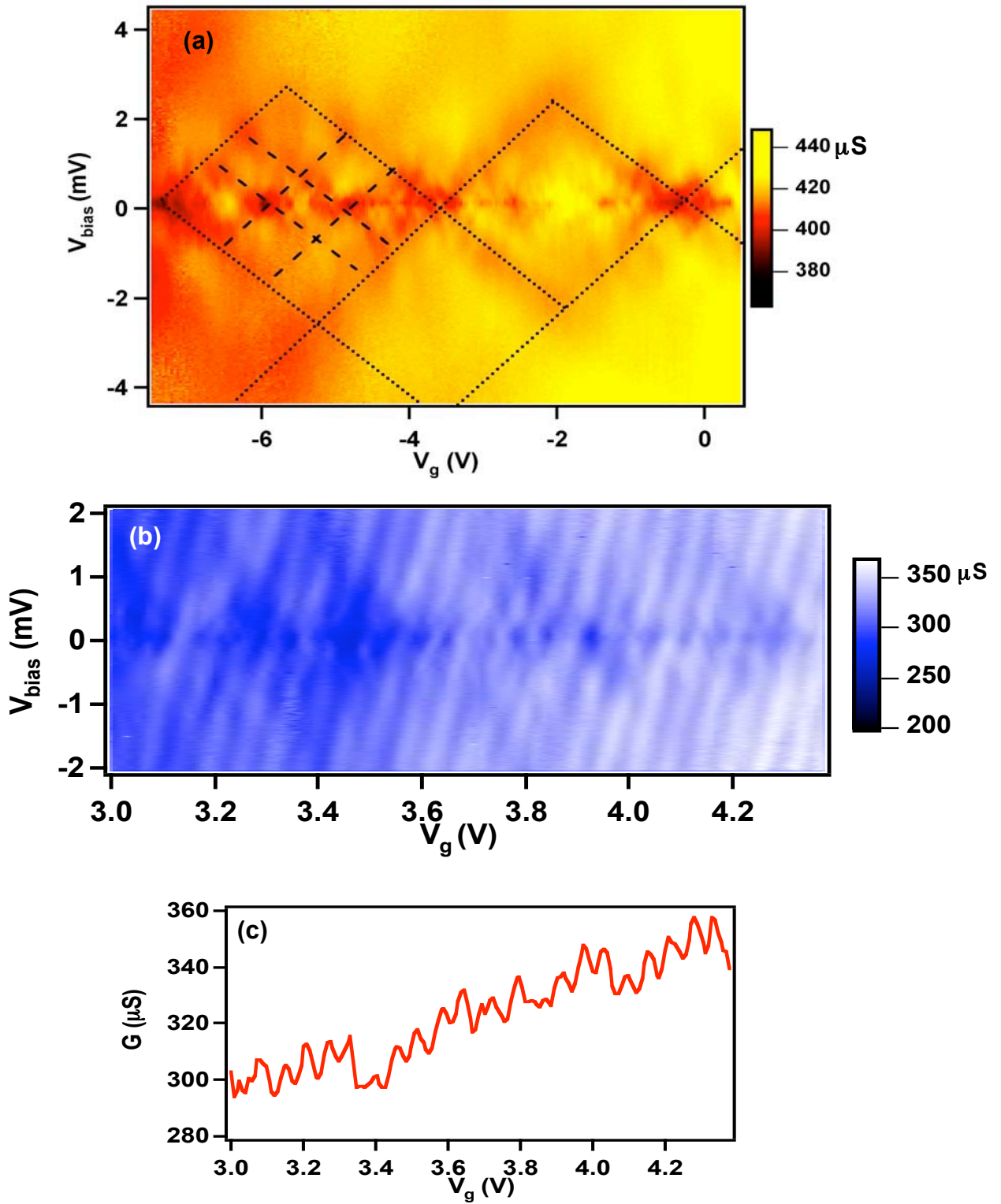

**Fig. 4.** (a) (b) Differential conductance (*dI/dV*) of Device 1 as functions of bias and gate voltage at B=0. The white band near zero bias are the enhanced conductance due to proximity effect. (b) *dI/dV* of Device 1 vs bias at *B*=0 (red curve) and B=50mT (blue curve). The data is taken at $V_g$=3.95 V, and the *B*=0 curve is offset by 20μS for clarity. Inset: *dI/dV* vs bias at $V_g$=3.32 V and *B*=0 with expanded bias scale. Notice the dip at ~2mV. (c) *dI/dV* of Device 7 vs bias at *B*=0 and *B*=50mT. $V_g$=-7.58V. (d) *dI/dV* of Device 7 as functions of *B* (horizontal axis) and source-drain bias (vertical axis). Color scale is same as 4e. The three red bands near zero bias correspond to enhanced conductance. (e) Same as 4d but with larger range of bias and *B*. The white and blue traces at higher bias correspond to the above-gap conductance dips. The dotted line is theoretical curve calculated using Eq. (3).

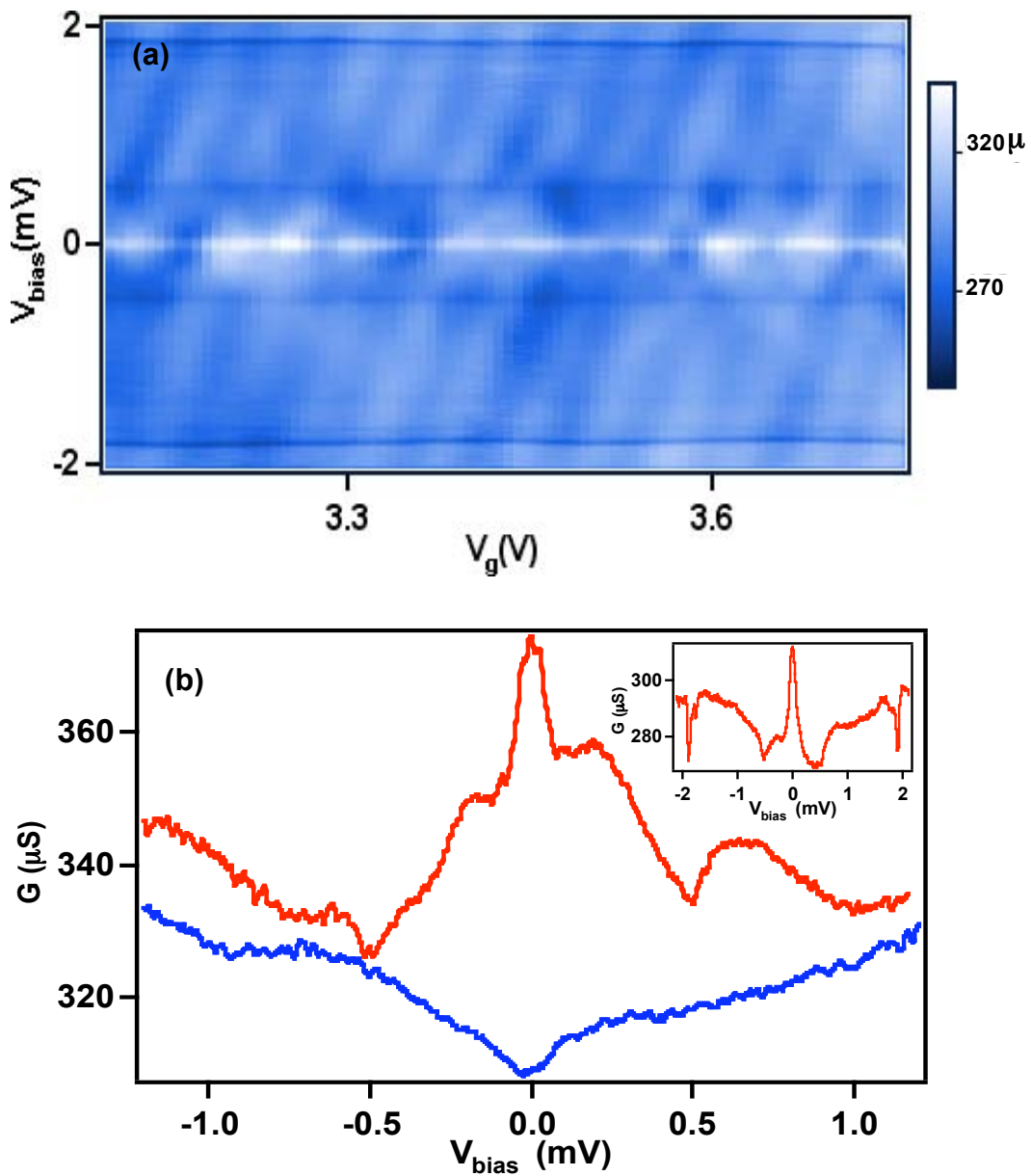

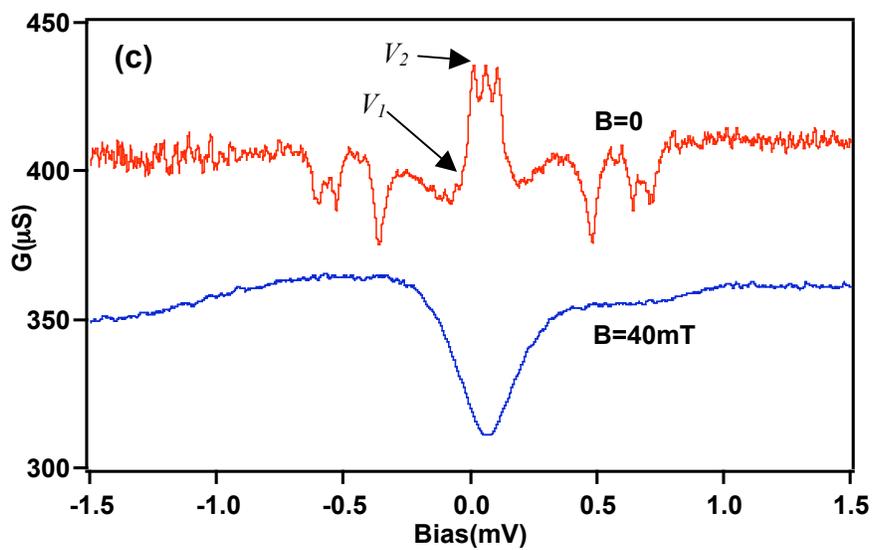
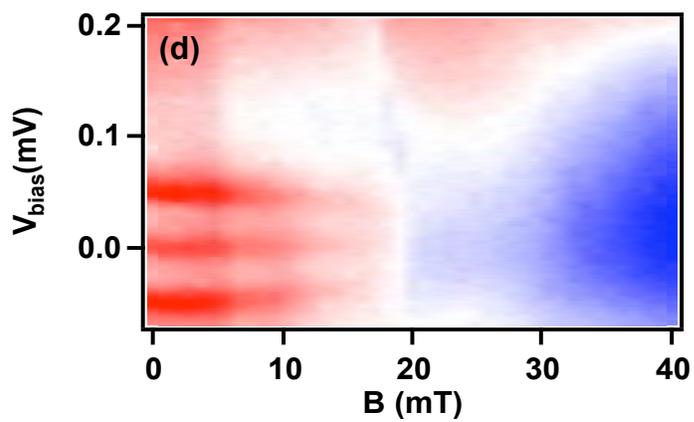
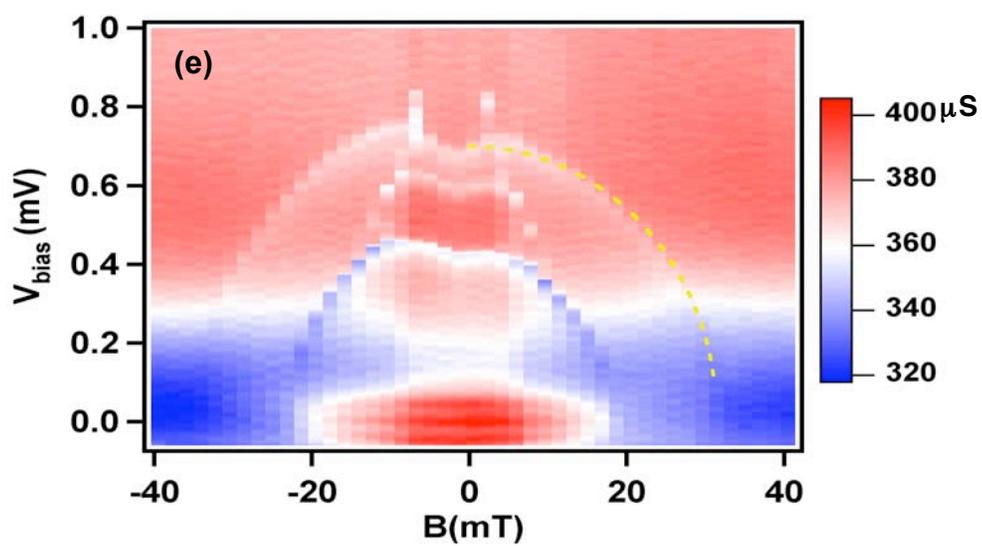